\documentclass{svjour2}                     
\smartqed  
\usepackage{graphicx}
\usepackage{natbib}
\usepackage{mathptmx}      
\journalname{Solar Physics}
\begin{document}

\title{PHOTOSPHERIC MAGNETIC FIELD OF THE SUN:
TWO PATTERNS OF THE LONGITUDINAL DISTRIBUTION}


\author{E.S.~VERNOVA  \and
        M.I.~TYASTO   \and
        D.G.~BARANOV
        }


\institute{E.S.~VERNOVA, M.I.~TYASTO \at
              IZMIRAN, SPb. Filial, St. Petersburg, Russia \\
              \email{helena@ev13934.spb.edu}
           \and
           D.G.~BARANOV \at
           A.F.~Ioffe Physical-Technical Institute,
                  St. Petersburg, Russia
}

\date{Received: date / Accepted: date}

\maketitle
\sloppy
\begin{abstract}
Longitudinal distributions of the photospheric magnetic field
studied on the basis of National Solar Observatory (Kitt Peak)
data (1976 -- 2003) displayed two opposite patterns during
different parts of the 11-year solar cycle. Heliolongitudinal
distributions differed for  the ascending phase and the maximum of
the solar cycle on the one hand, and for the descending phase and
the minimum on the other, depicting maxima around two
diametrically opposite Carrington longitudes ($180^\circ$ and
$0^\circ/360^\circ$). Thus the maximum of the distribution shifted
its position by $180^\circ$ with the transition from one
characteristic period to the other. Two characteristic periods
correspond to different situations occurring in the 22-year
magnetic cycle of the Sun, in the course of which both the global
magnetic field and the magnetic field of the leading sunspot in a
group change their sign. During the ascending phase and the
maximum (active longitude $180^\circ$) polarities of the global
magnetic field and those of the leading sunspots coincide, whereas
for the descending phase and the minimum (active longitude
$0^\circ/360^\circ$) the polarities are opposite. Thus the
observed change of active longitudes may be connected with the
polarity changes of Sun's magnetic field in the course of 22-year
magnetic cycle.

\keywords{Solar cycle \and Photospheric magnetic field
\and Longitudinal asymmetry}
\end{abstract}

\section{Introduction}
\label{intro} The 11-year cycle of solar activity reflects the
cyclic behavior of the Sun's large-scale magnetic field. In the
course of this cycle not only do the number and intensity of
various activity manifestations experience systematic changes but
so does their distribution over the solar surface . In contrast to
the well-established features of the latitudinal distribution of
the solar activity, seen for example in butterfly diagrams, there
exists a variety of evidence on the longitudinal distribution that
is difficult to reconcile in one unified picture.

The literature concerned with the problem of active longitudes is
too vast to give a detailed review here; we refer, therefore,  to
the review article by \citet{gaizauskas93}  and also to
\citet{bai03}, \citet{bigazzi04}, and references therein. Some
references concerning this problem are also given  in one of our
papers \citep{vernova04}. Here we shall confine ourselves to some
of the principal questions arising in the study of the active
longitudes.

An important problem is the stability and lifetime of the active
longitudes, which vary over a wide range in different studies.
Some authors have found active longitudes existing from several
successive solar rotations \citep{bumba65, detoma00} to 20 -- 40
consecutive rotations \citep{bumba69}. Yet according to
\citet{balthasar83} there is a "solar memory" for preferred
longitudes of activity extending at least over one, and probably
two cycles. Active zones persisting for three solar cycles  were
described  in \citet{bai88} and \citet{jetsu97}.

Another important problem is the localization of active longitudes
relative to the Carrington framework. Some authors observed that
the longitudes at which sunspots preferentially occur, do not
display differential rotation: The active longitudes experience
more rigid rotation with the Carrington period (T = 27.2753 days)
approximately. So, \citet{bumba69}  state that subsurface sources
that produce active longitudes rotate with the synodic period
regardless of the region latitude. Complexes of activity described
in \citet{gaizauskas83} were found to rotate at a steady rate,
sometimes coinciding with the Carrington rate and sometimes slower
or faster. The tendency of the activity of two successive solar
cycles to appear at the same preferred longitudes was found in
\citet{benevolenskaya99} and \citet {bumba00}.

In some  investigations constant rotation rate of active
longitudes with a period differing from the Carrington period was
registered. Two main periodicities (22.07 and 26.72 days) were
detected in the  major solar flare data \citep{jetsu97}. Moreover,
in some cases the rotation rate changed from cycle to cycle and
even during an individual solar cycle; see, {\it e.g.},
\citet{bogart82}.

In contrast these results favoring the rigid rotation of active
longitudes, \citet{berdyugina03} report two persistent active
longitudes $180^\circ$ apart, their migration relative to
Carrington frame being defined by the differential rotation and
the mean latitude of sunspot formation.

A number of investigations pointed out the tendency of the active
longitudes toward arrangement in the diametrically opposing
locations of the solar sphere  \citep{dodson68, bai87, jetsu97,
bumba00, mordvinov04}.

Some of the authors connect the longitudinal asymmetry with the
hypothesis of a relic magnetic field existing in the solar core.
It was pointed out that the appearance of some particular active
longitudes on the Sun, as well as other asymmetrical
characteristics of solar activity cycles, may be explained by the
presence of the relic magnetic field tilted with respect to the
solar rotation axis; see, {\it e.g.}, \citet{bravo95} and
\citet{kitchatinov01}.

Active longitudes were observed not only for sunspots, solar
flares, and photospheric magnetic fields but also for coronal mass
ejections \citep{skirgiello05}, in the solar wind and in the
interplanetary magnetic field \citep{neugebauer00}. Two long-lived
active longitudes $180^\circ$ apart were found on stars
\citep{jetsu93} with switching of activity ("flip-flopping")
between these two longitudes.

One can see that conclusions drawn by various authors differ from
each other essentially and sometimes they are even contradictory.
This can be explained, in our opinion, by the difference in the
considered time spans rather than by the different methods
employed. We have shown earlier \citep{vernova04, vernova05} that
heliolongitudinal distributions of various manifestations of solar
activity display two opposite patterns for the ascending phase and
the maximum of the 11-year solar cycle on the one hand and for the
descending phase and the minimum on the other. Longitudinal
distributions for these periods depicted maxima around two
opposite Carrington longitudes ($180^\circ$ and
$0^\circ/360^\circ$). Solar magnetic fields are known to be the
source of  many solar activity manifestations. Therefore it may be
of interest to  consider the spatial distribution of the
photospheric magnetic field  and its connection with  asymmetric
distribution of solar activity.

Sections~\ref{data} and~\ref{method} describe, respectively, the
data and the method used in this paper. In
Section~\ref{distribution} we examine some features of the
magnetic field longitudinal distribution. In
Section~\ref{polarity} the connection of solar active longitudes
with the magnetic field polarity is considered. In
Section~\ref{sunspots} the observed phenomenon is traced in
various manifestations of solar activity. In
Sections~\ref{discussion} and~\ref{conclusion} we discuss and
interpret the obtained results and draw our conclusions.
\section{Data}  
\label{data} For this study  synoptic maps of the photospheric
magnetic field produced at the National Solar Observatory/Kitt
Peak (available at http://nsokp.nso.edu/) were used. These data
cover period from 1975 to 2003 (Carrington rotations Nos.
1625~--~2006). As the data  have many gaps during initial period
of observation  we have included in our analysis data starting
from Carrington rotation No. 1646.

\begin{figure}
\begin{center}
\includegraphics{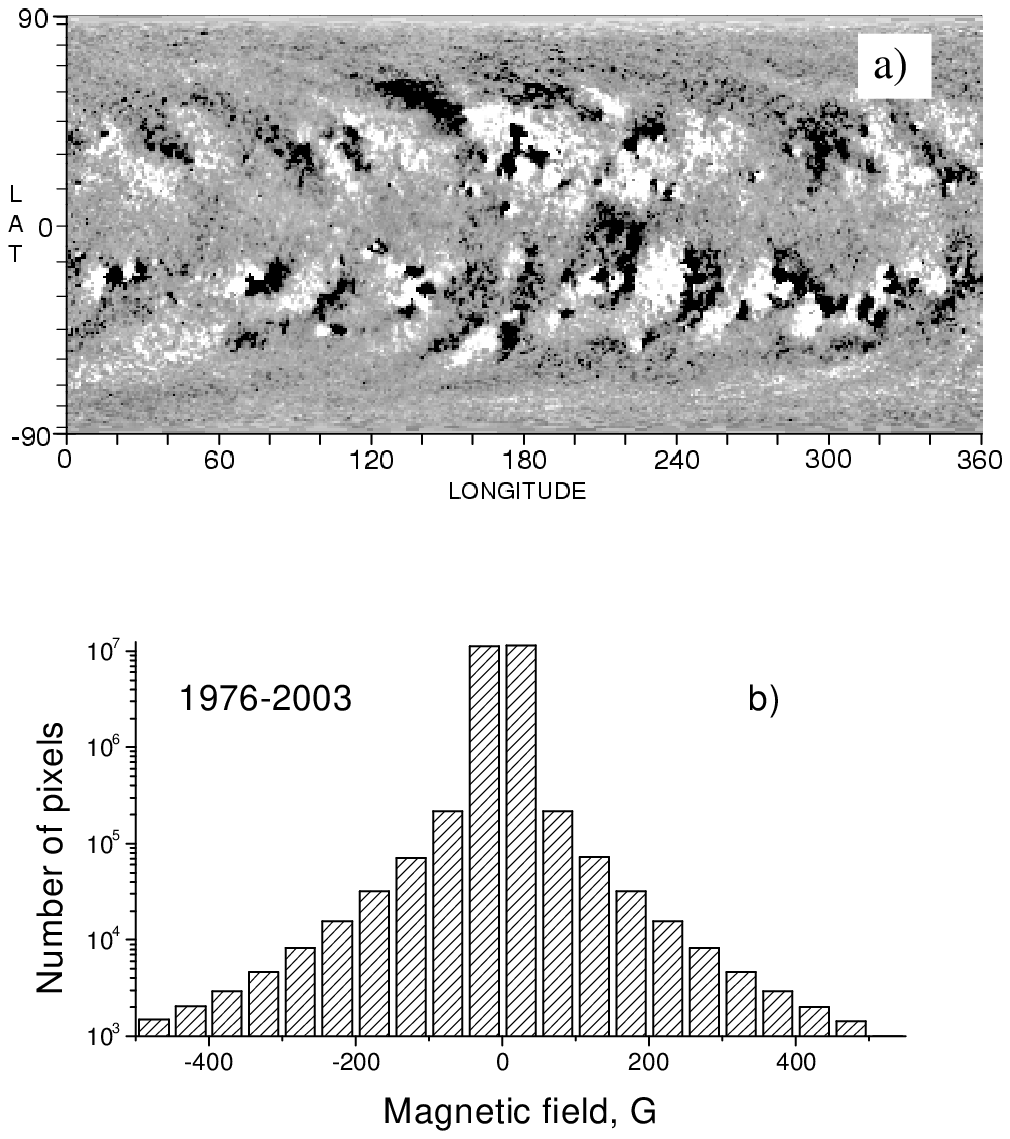}
\caption{NSO/Kitt Peak solar magnetic flux data. (a) Sample
synoptic map for  Carrington rotation No. 1689; white represents
positive polarity of magnetic field, and black represents negative
polarity. (b) Distribution of magnetic flux values for
1976~--~2003. (For better viewing the limited range of $\pm500$ G
is presented.)
     }
\label{synoptic}
\end{center}
\end{figure}

Synoptic maps have the following spatial resolution: $1^\circ$ in
longitude (360 steps); 180 equal steps in the sine of the latitude
from --1 (South pole) to +1 (North pole). Thus every map consists
of $360\times180$ pixels of magnetic flux values.

An example of a magnetic field synoptic map is presented in
Figure~\ref{synoptic}a (Carrington rotation No. 1689). White
represents the positive polarities of magnetic field and black the
negative polarities. One can see that strong magnetic fields of
both polarities occupy a relatively small part of the Sun's
surface. Most of the surface is covered by magnetic field of low
magnetic flux density (various shades of gray).

This  feature can be seen in Figure~\ref{synoptic}b, which depicts
the magnetic field strength distribution for the whole period
under consideration (1976~--~2003). A nearly symmetric
distribution contains 98.7\% of the values in the interval
0~--~100 G, whereas pixels with $B$ above 100 G occupy  only 1.3\%
of the solar surface. (For better viewing the limited interval
$|B|<500$ G is represented.) Even so, the total number of the
latter pixels is large enough, amounting to $3\times10^5$ for
1976~--~2003, and thus allowing detailed analysis of their
distribution.
\section{Method}
\label{method} 
\begin{figure}
\begin{center}
\includegraphics{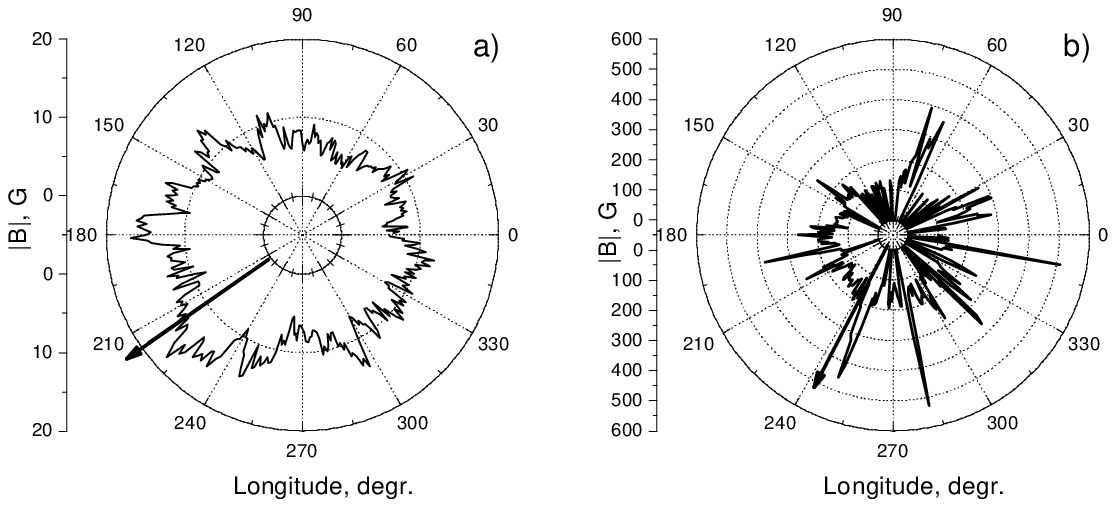}
\caption{Polar diagrams of latitude-averaged magnetic field for
(a) weak fields -- pixels with $|B| < 100$ G  and (b) strong
fields -- pixels with $|B| > 100$ G. Arrows represent longitudinal
asymmetry of the distribution evaluated by means of the vector
summation technique.  In the figure vector length is given in
relative units; direction points to the Carrington longitude
dominating during the given rotation. A latitude interval from
--90$^\circ$ to +90$^\circ$ was used in this example.
     }
\label{polar}
\end{center}
\end{figure}

To study the longitudinal distribution of the
photospheric magnetic field we have used the following approach.
For each $1^\circ$ interval of heliolongitude from $1^\circ$ to
$360^\circ$, magnetic flux values were averaged over the selected
latitude interval. Latitude-averaged  values of $|B|$ were used to
study the longitudinal magnetic field  distribution. An example of
the resulting longitudinal distribution for Carrington rotation
No. 1689 is presented in Figure~\ref{polar}. The whole latitude
interval (from --90$^\circ$ to +90$^\circ$) was used in this
example. Two cases were considered separately:  weak fields
(pixels with $|B|<100$ G, Figure~\ref{polar}a) and strong fields
($|B|>100$ G, Figure~\ref{polar}b).

The longitudinal asymmetry of the photospheric magnetic field
distribution was evaluated by means of the vector summation
technique \citep{vernov79, vernova02}. Latitude-averaged values of
$|\mathbf{b}_i|$ for each $1^\circ$ interval (from $1^\circ$ to
$360^\circ$) were considered as vector modulus, with corresponding
Carrington longitude being the phase angle of $\mathbf{b}_i$. Then
resulting vector sum was calculated for every Carrington rotation
as
$$
\mathbf{B} = \sum\limits_{i=1}^{360}\mathbf{b}_i.
$$
Whereas the modulus of  $\mathbf{B}$ (of the resulting vector sum)
can be considered as a  measure of longitudinal asymmetry, the
direction of the vector (phase angle) points to the Carrington
longitude dominating during the given Carrington rotation.
Long-term change of the longitudinal asymmetry as well as that of
the total photospheric magnetic flux follow the 11-year solar
cycle.

Vectors presented as arrows in Figure~\ref{polar}  show both
values of the longitudinal asymmetry (vector modulus in relative
units) and Carrington longitudes dominating in this rotation
(vector phase angle). Two cases -- weak fields (pixels with
$|B|<100$ G, Figure~\ref{polar}a) and strong fields ($|B|>100$ G,
Figure~\ref{polar}b) -- were treated separately. The longitudinal
distribution for the strong field contains distinct separate
peaks, whereas the weaker field displays smoother longitude
dependence. However, in spite of the visible difference in the
general pattern, the resulting vectors of the longitudinal
asymmetry have fairly close orientations in this example.

By means of  this technique the dominating Carrington longitude
was evaluated for each solar rotation.  The longitudinal
distribution of the photospheric magnetic field during
1976~--~2003 was studied on the basis of the time series obtained
in this way. Only strong magnetic fields ($|B|>100$ G) were
included further in our calculations. The latitude interval from
--45$^\circ$ to +45$^\circ$ is considered throughout the rest of
this paper. In addition to the combined distribution for both
hemispheres data for the northern hemisphere (latitudes from
$0^\circ$ to $+45^\circ$) and for the southern one (latitudes from
$0^\circ$ to $-45^\circ$) were considered separately.
\section{Two Patterns of Magnetic Field Longitudinal Distribution}
\label{distribution} While studying the sunspot longitudinal
distributions corresponding to the four phases of the solar cycle
(minimum, ascending phase, maximum, and descending phase), we have
found considerable similarity between ascending phase and maximum
\citep{vernova04}. The descending phase and minimum distributions
on the other hand displayed a pattern drastically different from
the one typical for two other phases of the solar cycle.
Accordingly, we grouped the whole data set into two parts, one (to
be called AM) consisting of the ascending phases and the maxima of
the solar cycles included in the data set and the other (DM)
consisting of the descending phases and the minima.

\begin{figure}
\begin{center}
\includegraphics{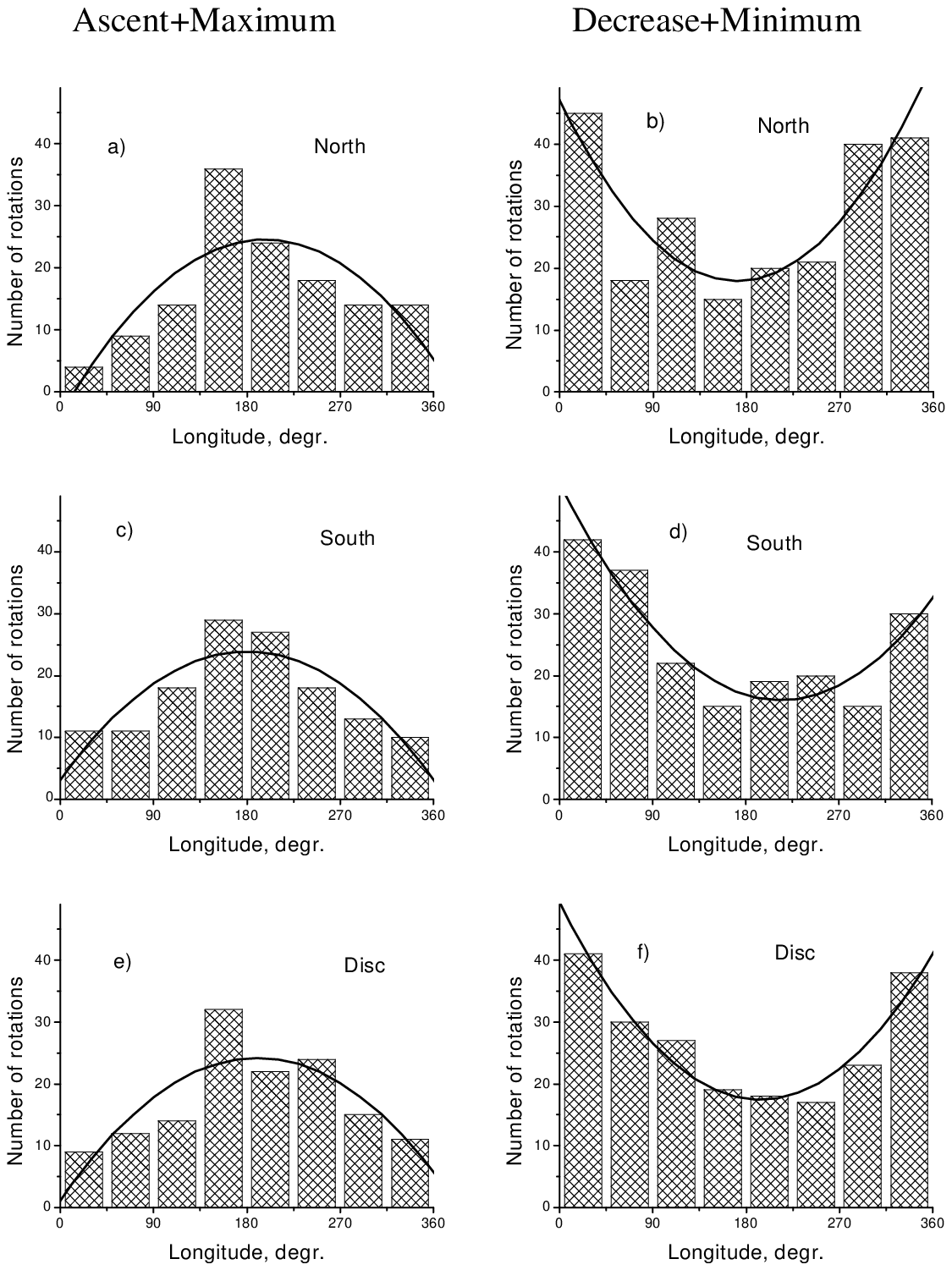}
\caption{Two opposite patterns  of magnetic field longitudinal
distribution corresponding to  the ascent and the maximum of solar
cycle  (left panels) and to the decrease and the minimum (right
panels): (a), (b) northern hemisphere; (c), (d) southern
hemisphere; (e), (f)  solar disk. Histograms represent
distributions of dominating longitudes (one value per Carrington
rotation). Solid lines indicate the best-fitting second-order
polynomial, drawn to emphasize the difference between two forms of
envelope. Note that histograms for solar disk [(e) and (f)] do not
represent the sum of the northern and southern hemisphere
histograms. They were obtained by independently treating of the
data  from --45$^\circ$ to +45$^\circ$ in heliolatitude.
     }
\label{magnetic}
\end{center}
\end{figure}
The same approach is used now in the study of the photospheric
magnetic field distribution. Longitudinal distributions calculated
by means of the technique described  in Section~\ref{method} are
presented in Figures~\ref{magnetic}a and b (northern hemisphere,
latitudes  from $0^\circ$ to $+45^\circ$), Figures~\ref{magnetic}c
and d (southern hemisphere, latitudes from $0^\circ$ to
$-45^\circ$), and Figures~\ref{magnetic}e and f (solar disk,
latitudes from $-45^\circ$ to $+45^\circ$). These figures
represent distributions of dominating longitudes (one value per
Carrington rotation) for the following time intervals. All AM
phases for the 1976~--~2003 period are combined in
Figures~\ref{magnetic}a, c, and e and all DM phases are combined
in Figures~\ref{magnetic}b, d, and f. We have taken into account
only those Carrington rotations for which the asymmetry of the
distribution was pronounced ({\it i.e.}, the modulus of the
resulting vector $\mathbf{B}$ was sufficiently large). This has
reduced the sample only by about 10\%.

Two opposite types of photospheric field longitudinal distribution
can be seen for two parts of the 11-year cycle: convex -- for the
ascending phase and maximum (Figures~\ref{magnetic}a, c, and e)
and concave -- for the decrease and minimum
(Figures~\ref{magnetic}b, d, and f). Accordingly the distribution
maxima change from $180^\circ$ for the AM phase to
$0^\circ/360^\circ$ for the DM phase. The best-fitting
second-order polynomials were added in each case to emphasize the
convex (maximum close to $180^\circ$ of Carrington longitude) or
concave (maximum close to $0^\circ/360^\circ$ longitude) nature of
the distribution. One can see in Figure~\ref{magnetic} that two
patterns of the longitudinal distribution appear in the northern
and the southern hemispheres simultaneously indicating a
synchronous development of the asymmetry in both solar
hemispheres. Yet the distribution for the northern hemisphere
depicts a less regular behavior than that for the southern
hemisphere.

As was stated in Section~\ref{method}, only strong magnetic fields
are considered in this study. The chosen limit ($|B|>100$ G) makes
it possible to observe the indicated phenomenon (two maxima
separated by  $180^\circ$) more clearly. The same effect can be
seen, though it is less pronounced, when we lower the limit to
$|B|>10 - 20$ G. It should be noted, however,  that for weak
fields (with $|B|<20$ G) the effect disappears completely; the
longitudinal distribution is uniform within the range of
statistical error.

\begin{figure}
\begin{center}
\includegraphics{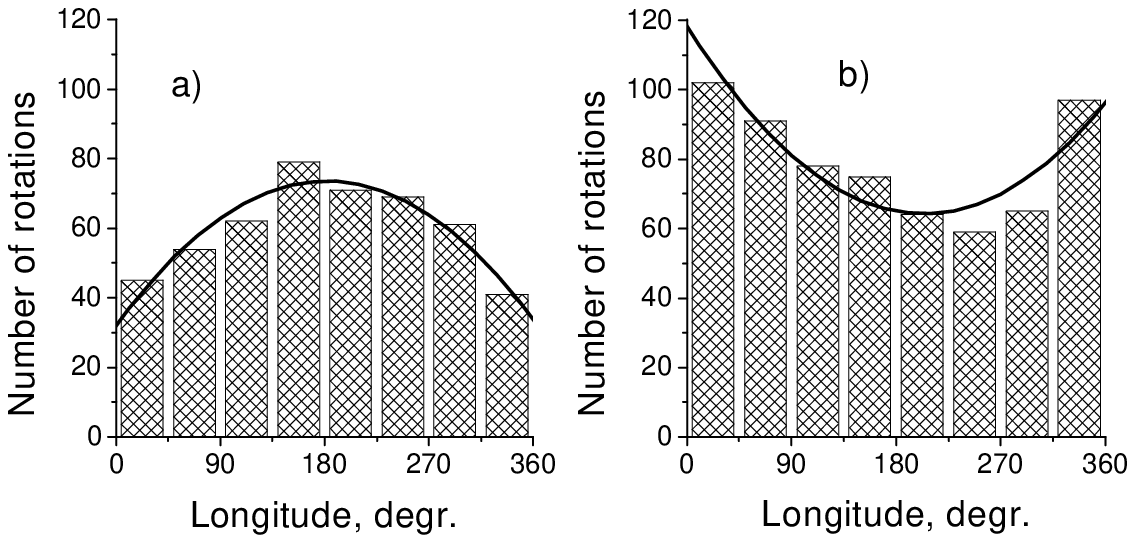}
\caption{Sunspot longitudinal distribution (solar disk,
1917~--~2003): (a) the ascent and the maximum (AM) of solar cycle;
(b) the decrease and the minimum (DM). Solid lines indicate the
best-fitting second-order polynomials. Two opposite patterns can
be seen for the AM and DM periods of the solar cycle.
     }
\label{spots}
\end{center}
\end{figure}

The distributions in Figure~\ref{magnetic} were constructed as a
combination of data for several solar cycles, ({\it i.e.}, the
observed peculiarities point to long-lived longitudinal asymmetry
distinguishable through averaging of long data sets).
  Our previous results \citep{vernova04} obtained for the
longitudinal sunspot distribution show that these distribution
peculiarities go beyond the period 1976~--~2003. For the present
article we treated anew the data set produced by the Greenwich
observatory (http://solarscience.msfc.nasa.gov/greenwch.shtml)
using the same method of vector summing (see Section~\ref{method})
for the period 1917~--~2003. Each sunspot was represented as a
vector: The vector modulus was set equal to the sunspot area and
the direction of the vector (phase angle) was pointed to the
Carrington longitude of the sunspot. As a result of vector
summing, the dominating longitude for each solar rotation was
obtained. In Figure~\ref{spots} the longitudinal distributions of
sunspots for AM and DM periods are presented. Here, as in the case
of magnetic fields, the same patterns of the distribution can be
seen (convex for the AM periods and concave for the DM ones).
Thus, we see that, when the data are averaged over 8 solar cycles,
the form of the distribution is preserved with the shift of the
maximum by $180^\circ$ when passing from the AM period to the DM
one.

\begin{figure}
\begin{center}
\includegraphics{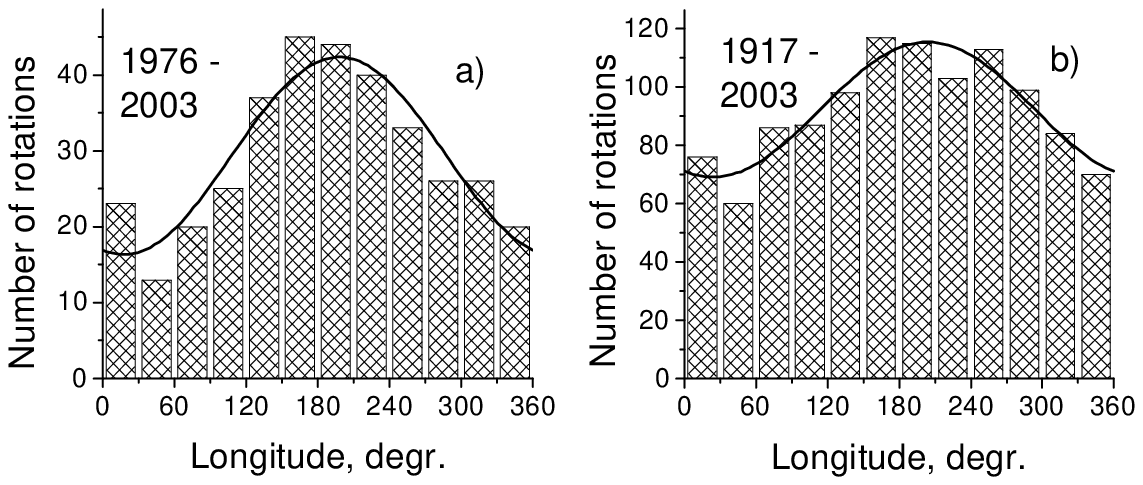}
\caption{Combination of  all data in a single distribution. The
distribution for the ascent--maximum of solar cycle was combined
with the decrease--minimum distribution shifted by $180^\circ$.
(a) solar magnetic field, 1976~--~2003 (latitudes from
--45$^\circ$ to +45$^\circ$), and (b)  sunspots, 1917~--~2003.
Solid lines are the best-fitting sine curves.
     }
\label{shifted}
\end{center}
\end{figure}

 That we have
to deal with two parts  of the same process can be illustrated in
the following way. A distribution combining all magnetic field
data for 1976~--~2003 was obtained by shifting the
decrease--minimum data by $180^\circ$ and summing with the
ascent--maximum distribution (Figure~\ref{shifted}a). In the same
way sunspot data for 1917~--~2003 were combined in a single
distribution both for AM and DM periods (Figure~\ref{shifted}b).

The resulting modified histograms depict a very systematic (almost
sinusoidal) behavior of the longitudinal distribution with a
pronounced maximum at about $180^\circ$. In spite of long data
sets under consideration,  the resulting histograms display
extraordinary large and smooth variations. Best-fitting sinusoids
were added to Figures~\ref{shifted}a and b. The maximum of the
fitting sinusoid is located at $198\pm7^\circ$ longitude for the
magnetic field  (Figure~\ref{shifted}a) and at $204\pm7^\circ$
longitude for the sunspot distribution (Figure~\ref{shifted}b). A
chi-square test shows that the probability for the pattern
appearing from random fluctuations of the uniform distribution is
much less than 1\% for both distributions (Figures~\ref{shifted}a
and b).

The present phenomenon  manifests itself as a result of averaging
over the period of 28 years for magnetic field and 87 years for
sunspots. It shows the alternating domination of two large
intervals of the longitudes: from $90^\circ$ to $270^\circ$ (with
the maximum at $180^\circ$) for the AM period, and the opposite
interval from $270^\circ$ to $90^\circ$ (the maximum at
$0^\circ/360^\circ$) for the DM period. This effect does not
exclude the clustering of solar activity in one, two, or more
active zones, whose locations can be different from the maxima of
the longitudinal asymmetry observed by us.
\section{Magnetic Field Polarity and Active Longitudes}
\label{polarity} The times separating the two characteristic
periods (AM and DM) are important intervals of the solar magnetic
cycle. The time between the solar maximum and the beginning of the
declining phase coincides with the inversion of the Sun's global
magnetic field whereas the time between the solar minimum and the
ascending phase is related to the start of the new solar cycle and
the change of the magnetic polarity of sunspots according to
Hale's law.
\begin{figure}
\begin{center}
\includegraphics{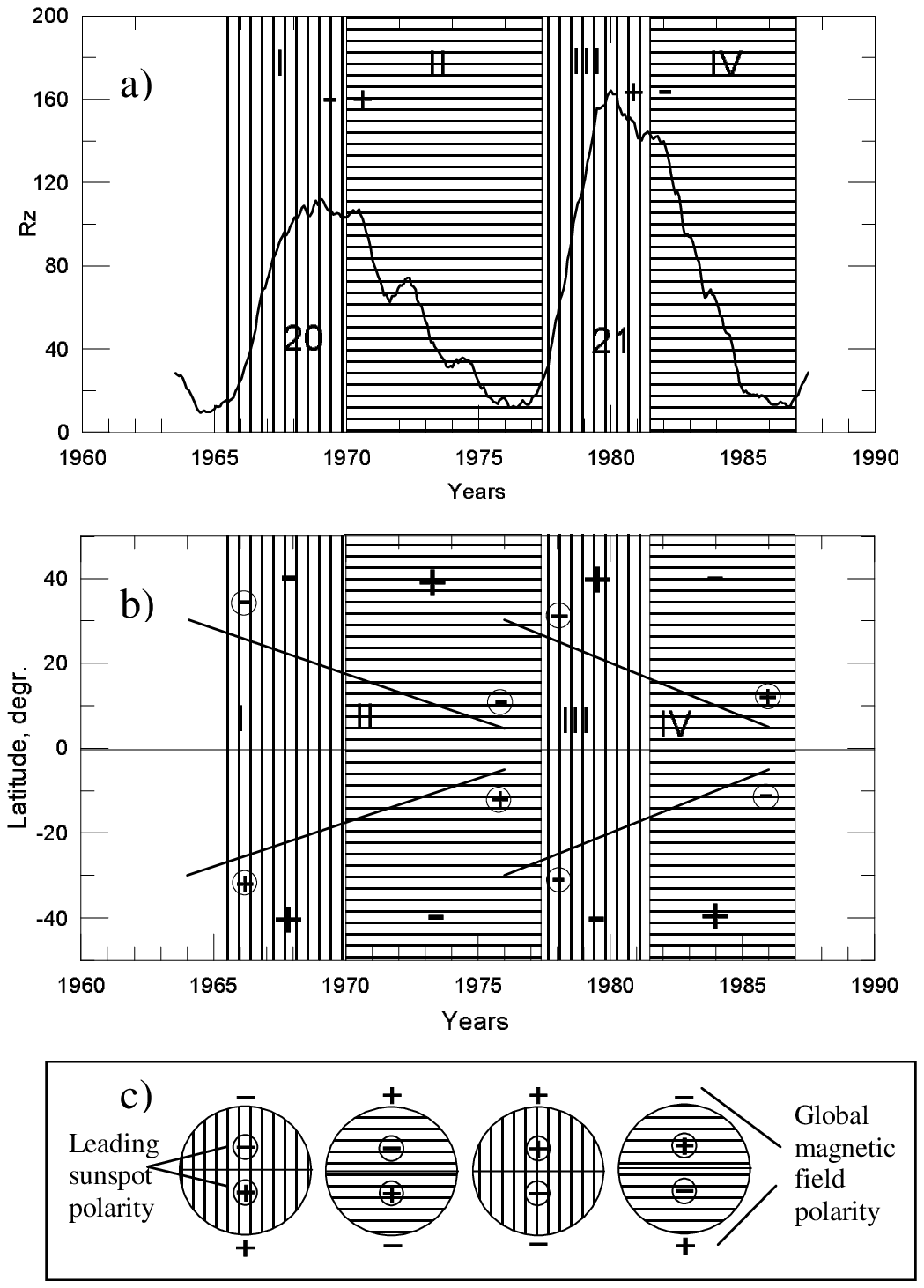}
\caption{Change of global and local magnetic field polarities in
the course of the solar magnetic cycle. (a) Example of  two
11-year cycles (20th and 21st) forming a single magnetic cycle:
solid line -- Wolf number; I and III -- ascent and maximum periods
(vertical shading), II and IV -- decrease and minimum (horizontal
shading). Global magnetic field polarities are designated at the
top of the panel for the North hemisphere. (b) Evolution of local
magnetic fields during the same period. Solid lines show
schematically the latitudinal drift of sunspots. Leading sunspot
polarities are shown inside  circles. Other notation is the same
as in the panel (a). Global magnetic field polarities are
designated at the top and at the bottom of the panel. (c) Scheme
illustrating coincidence of global magnetic field polarity with
the polarity of leading sunspots for periods I and III (ascent and
maximum). Opposite polarities occur for periods II and IV
(decrease and minimum).
     }
\label{cycle}
\end{center}
\end{figure}

The change of magnetic field polarities in the course of the
22-year solar cycle is illustrated by Figure~\ref{cycle}, where
two successive 11-year cycles are presented (by way of example,
even cycle 20 and odd cycle 21 are considered). According to our
approach each of the cycles is divided into two periods
(ascent--maximum and decrease--minimum). Four corresponding
periods I, II, III, and IV are separated by vertical lines
(Figure~\ref{cycle}a).

The polarity changes of both the global  and the local solar
magnetic fields are shown schematically in Figure~\ref{cycle}b.
The polarity of the leading sunspot is indicated inside the
circles at the beginning and at the end of lines following the
latitudinal displacement of the sunspot activity during the solar
cycle. At the top and at the bottom of the panel the global
magnetic field polarities are indicated for the northern and
southern hemispheres, respectively.

Connection of the four periods I, II, III, and IV with the
magnetic field polarities is depicted in Figure~\ref{cycle}c. Big
circles represent the Sun during the four periods, with global
magnetic field polarity being indicated at the poles. Small
circles represent schematically the leading sunspots and their
polarity for the corresponding solar hemisphere.

Comparing the four diagrams in Figure~\ref{cycle}c one can see
that the two characteristic periods (AM and DM) correspond to
different situations occurring in the 22-year magnetic cycle of
the Sun, in the course of which the global magnetic field polarity
and polarity of the leading sunspot may be the same (for each
solar hemisphere) or opposite. During the ascending phase and the
maximum (active longitude $180^\circ$) polarities of the global
magnetic field and those of the leading sunspots coincide, whereas
for the descending phase and the minimum (active longitude
$0^\circ/360^\circ$) polarities are opposite.

\begin{table}[h]
\caption{Solar magnetic field polarities and active longitudes}
\label{mf2}
\begin{center}
\begin{tabular}{cccccc}
\hline\noalign{\smallskip} &\multicolumn{2}{c}{North
hemisphere}&\multicolumn{2}{c}{South hemisphere}&
 \\
 \noalign{\smallskip}\cline{2-5}
 \\
Solar cycle & Global   & Leading   & Global   & Leading  & Active  \\
 phase & magnetic & sunspot   & magnetic & sunspot  & longitude\\
       & field    & polarity  & field    & polarity &         \\
\noalign{\smallskip}\hline\noalign{\smallskip}
\multicolumn{6}{c}{Even  cycle}\\
\noalign{\smallskip}\hline\noalign{\smallskip}
Ascent--maximum   & $-$ & $-$ & $+$ & $+$ & $180^\circ $ \\
Decrease--minimum  & $+$ & $-$ & $-$ & $+$ & $ 0^\circ  $ \\
\noalign{\smallskip}\hline\noalign{\smallskip}
\multicolumn{6}{c}{Odd  cycle}\\
\noalign{\smallskip}\hline\noalign{\smallskip}
Ascent--maximum   & $+$ & $+$ & $-$ & $-$ & $180^\circ $ \\
Decrease--minimum & $-$ & $+$ & $+$ & $-$ & $ 0^\circ  $ \\
\noalign{\smallskip}\hline
\end{tabular}
\end{center}
\end{table}

In Table 1 the polarity of the global magnetic field, polarity of
the leading sunspot and longitudinal location of the magnetic
field maximum are compared for periods I, II, III, and IV of the
solar magnetic cycle. One can see that the maximum of the
photospheric magnetic field distribution appeared at longitude
$180^\circ$ during AM periods (I and III) in contrast to DM
periods (II and IV) with the maxima at $0^\circ/360^\circ$.

Thus when the polarity of the global magnetic field and the
polarity of the leading sunspot are the same (within one
hemisphere) the longitudinal maximum is at $180^\circ$. However,
when these polarities are opposite the maximum is at about
$0^\circ/360^\circ$. One may suppose that the observed change of
active longitudes is connected with the polarity changes of Sun's
magnetic field in the course of the 22-year magnetic cycle.
\section{Longitudinal Distributions of Solar Activity}
\label{sunspots} Analogous results were obtained by us earlier on
the basis of Wilcox Solar Observatory (WSO) data for 1976~--~2004
\citep{vernova05}. For treating the WSO photospheric magnetic
field data another technique was used instead of the vector
summation used here for the Kitt Peak data. Latitude-averaged
magnetic field values (latitudes from $-70^\circ$ to $+70^\circ$)
for each solar rotation were analyzed and peaks exceeding a mean
value by 1.5 standard deviations were selected. Longitudes
corresponding to the positions of these peaks were used when
plotting the magnetic field longitudinal distribution. In contrast
to Kitt Peak data, no discrimination was made between strong and
weak fields; that is, all measured values ($30\times72$ pixels for
each synoptic map) were included.

For both data sets two opposite types of longitudinal
distributions corresponding to AM and DM periods were observed for
the photospheric magnetic field, with maxima being situated at
Carrington longitude $180^\circ$ during the AM period and at
$0^\circ/360^\circ$ during the DM one.

Longitudinal distributions of many different manifestations of
solar activity display the same features as those that are
characteristic of the photospheric magnetic field distribution and
sunspots. In the study of solar proton event sources
(1976~--~2003) and X-ray flare sources (1976~--~2003) it was found
that activity distributions behave differently during the
ascending phase and the maximum of the solar cycle on the one hand
and during the descending phase and the minimum on the other,
depicting maxima around opposite Carrington longitudes
($180^\circ$ and $0^\circ/360^\circ$) \citep{vernova05}. Thus
various manifestations of solar activity (sunspots, X-ray flares,
and solar proton event sources), as well as the photospheric
magnetic field  show long-term asymmetry with similar features of
heliolongitudinal distribution.

This phenomenon is seen more clearly when the vector summation
method is used for data treatment as it was in cases of sunspots,
Kitt Peak magnetic field data, and X-ray flare sources. Yet
similar results were obtained in our studies with other kinds of
treatment (sunspots, WSO magnetic field data, and solar proton
event sources). It should be noted that, along with the phenomenon
found by us, other types of longitudinal asymmetry may exist, and
these could manifest themselves, for example, during shorter
periods of time.
\section{Discussion}
\label{discussion} The effects described here evidence the
existence of a specific longitudinal asymmetry of the solar
photospheric magnetic field distribution. On the one hand, this
asymmetry is very stable; the maxima of the magnetic field
distribution are observed at the same Carrington longitudes over a
period of 1976~--~2003 (with the same feature persisting for
1917~--~2003 in the sunspot distribution). On the other hand, the
maximum of distribution changes jump like its longitudinal
position by $180^\circ$ twice during the 11-year solar cycle. The
long-term stability of the distribution relative to Carrington
coordinate frame can be interpreted as a manifestation of rigidly
rotating magnetic field structure.

Of special interest is the regular change of the maximum position
(around $180^\circ$ for the ascent and maximum period and
$0^\circ/360^\circ$ for the decrease and minimum period)  found by
us earlier for the sunspot distribution \citep{vernova04} and for
other indices of solar activity \citep{vernova05} and now
confirmed by the analysis of the photospheric magnetic field data.
The connection of the two characteristic periods  with changes of
the global magnetic field polarity and those of the leading
sunspot polarities is unlikely to be a coincidence. It was shown
by \citet{benevolenskaya02} that the relation of the polarity of
the Sun's global magnetic field to the sign of the following
sunspots is of principal importance. Using soft X-ray data from
the Yohkoh X-ray telescope for the period of 1991~--~2001 they
discovered giant magnetic loops connecting the magnetic flux of
the following parts of the active regions with the magnetic flux
of the polar regions that have the opposite polarity. These
large-scale magnetic connections appeared mostly during the rising
phase of the solar cycle and its maximum. These connections did
not appear or were very weak during the declining phase of the
solar cycle. Thus the two characteristic periods AM and DM
correspond to radically different patterns of the large-scale
magnetic field of the Sun.

The features of the magnetic field distribution described here are
in good accordance with those previously reported by us for the
solar activity distribution. This coincidence points to the
asymmetry of the solar magnetic field distribution being the cause
of the corresponding longitudinal asymmetry of various
manifestations of the solar activity. It should be noted that all
cases displayed a broad maximum of the longitudinal distribution,
which does not conform with  the idea of active longitudes having
an extent of about $20^\circ$~--~$60^\circ$, see, {\it e.g.},
\citet{benevolenskaya99}. It is possible that some smoothing of
the distribution can be attributed to averaging over the long time
interval. Yet other authors have also pointed out a similar
pattern of the longitudinal asymmetry. \citet{balthasar83}
observed the longitudinal asymmetry of sunspot groups as the
predominance of one of the solar hemispheres as compared with the
other one.

Both for solar activity manifestations  \citep{vernova04} and for
the photospheric magnetic field (this study) the maxima of
distributions were observed near the heliolongitudes $180^\circ$
and $0^\circ/360^\circ$. It was noted by many authors that active
longitudes tend to appear on diametrically opposite sides of the
Sun.

Analysis of the solar activity for 1962~--~1966 has shown that
centers of activity have a tendency to develop on opposite sides
of the Sun \citep{dodson68}. In studies of the longitudinal
distribution of the major solar flares, active zones separated by
$180^\circ$ were found, where the flare occurrence rate was much
higher \citep{bai87}.

In some cases increased manifestations of activity were found near
the longitudes $180^\circ$ and $0^\circ/360^\circ$
\citep{dodson68, bumba96, jetsu97}. In \citet{jetsu97} a search
for periodicity was performed in major solar flare data during
three decades (1956~--~1985). For combined data of both
hemispheres two active longitudes rotating with a constant synodic
period of 22.07 days were observed at about $0^\circ$  and
$180^\circ$. One should especially mention the work of
\citet{skirgiello05}, where increased activity of heliolongitudes
$180^\circ$ and $30^\circ$ was discovered on the  basis of the CME
data; moreover, these nearly antipodal longitudes dominated
alternately. Similar to our observations, the domination of the
longitude $180^\circ$ coincided with the AM phase of the solar
cycle, whereas the domination of the longitude $30^\circ$
coincided with the DM phase. In agreement with our study of the
sunspot distribution \citep{vernova04}, the author points out for
the CME distribution a more distinct manifestation of the active
longitudes for the southern hemisphere of the Sun.

The transition from  the AM period to  the DM one, which coincides
with the inversion of Sun's global magnetic field, is accompanied
by radical changes in longitudinal distribution of the solar
activity. Simultaneously with the change of the activity maximum
position by $180^\circ$, the relative value of the asymmetry of
the northern and the southern hemispheres changes too. On the
basis of the sunspot data we have shown that during the period of
the inversion a transition from domination of the northern
hemisphere to domination of the southern hemisphere occurs
\citep{vernova02}. Similar features were found by \citet{kane05}
when studying the solar flare index: During cycles 21 and 22 the
transition from a multiyear northern preference to a multiyear
southern preference was observed around or after the polarity
reversal.
\section{Conclusion}
\label{conclusion} Longitudinal distributions of the photospheric
magnetic field studied on the basis of National Solar Observatory
(Kitt Peak) data (1976~--~2003) displayed two opposite patterns
during different parts of the 11-year solar cycle.
Heliolongitudinal distributions differed for  the ascending phase
and the maximum of the solar cycle on the one hand, and for the
descending phase and the minimum on the other, depicting maxima
around two opposite Carrington longitudes ($180^\circ$ and
$0^\circ/360^\circ$). Thus the maximum of the distribution shifted
its position by $180^\circ$ with the transition from one
characteristic period to the other.

The same peculiarities of the longitudinal distribution were
observed for the photospheric magnetic field studied on the basis
of Wilcox Solar Observatory data (1976~--~2004), sunspot data
(1917~--~2003); solar proton event sources (1976~--~2003), and
X-ray flare sources (1976--2003).

Two characteristic periods correspond to different situations
occurring in the 22-year magnetic cycle of the Sun, in the course
of which the global magnetic field polarity and polarity of the
leading sunspot may be the same (in each solar hemisphere) or
opposite. During the ascending phase and the maximum (active
longitude $180^\circ$) polarities of the global magnetic field and
those of the leading sunspots coincide, whereas for the descending
phase and the minimum (active longitude  $0^\circ/360^\circ$) the
polarities are opposite. Thus radical change of longitudinal
distribution may be connected with reorganization of solar
magnetic field in the course of the 22-year magnetic cycle.

The "classical" solar dynamo model is axisymmetric and therefore
does not explain the longitudinal asymmetry of the solar activity
distribution. The longitudinal asymmetry does not show itself in
such a regular and evident form as the latitudinal development of
the solar activity. That is why, up to a certain point, models of
the solar cycle simply ignored this aspect of the solar activity
distribution. Now the bulk of observational data confirming the
existence of active longitudes is so large that this phenomenon
should be accounted for in  a complete model of the solar cycle.
Efforts were made to explain the appearance of modes in the solar
cycle that are non-symmetric with respect to the axis \citep[
e.g.,][]{gilman97, dikpati05}. Peter Gilman in his talk "A 42 Year
Quest to Understand the Solar Dynamo and Predict Solar Cycles"
\citep{gilman06} expressed hope that "a unified theory of the
solar cycle and active longitudes" will be produced in the near
future.

\begin{acknowledgements}
NSO/Kitt Peak data used here are produced cooperatively by
NSF/NOAO, NASA/GSFC, and NOAA/SEL. Sunspot distribution was
studied on the basis of  Royal Greenwich Observatory/USAF/NOAA
Sunspot Record produced by Dr. David H. Hathaway.
\end{acknowledgements}


\end{document}